\RequirePackage{fix-cm}
\documentclass[smallextended]{svjour3}       
\smartqed  
\usepackage{graphicx}
\usepackage{comment}
\usepackage{color}
\begin{document}

\title{Status of the search for $\eta$-mesic nuclei with particular focus on $\eta$-Helium bound states
}


\author{Magdalena Skurzok}

\institute{M. Skurzok \at
              Institute of Physics, Jagiellonian University, prof. Stanisława {\L}ojasiewicza str. 11, 30-348 Krak\'ow, Poland \\
              Tel.: +48 12 664 4589\\
              \email{magdalena.skurzok@uj.edu.pl}\\           
              \emph{Present address:} INFN, Laboratori Nazionali di Frascati, 00044 Frascati, Italy
}

\date{Received: date / Accepted: date}

\maketitle

\begin{abstract}
In this paper the search for $\eta$-mesic nuclei with particular focus on light $\eta$-He bound states is reviewed. 
A brief description of recent experimental results is presented. 


\end{abstract}

\section{Introduction}~\label{Sec.1}

The possible existence of mesic nuclei, being bound states of a nucleus and a neutral meson ($\eta$, $\eta'$, K or $\omega$), is 
presently a prime topic of investigation 
in subatomic physics,
%
both from the theoretical
and experimental
standpoints;
for recent reviews, see \cite{KelkarRPP2013,MetagPPNP2017,BassRMP2019}.
%
%
One of the most promising candidates for these purely strong  interaction objects are  $\eta(\eta')$-mesic nuclei. This follows from the fact that the $\eta$-nucleon 
and 
$\eta'$-nucleon
interactions 
are reported to be attractive~\cite{BhaleraoPRL1985,HaiderPLB1986}, 
with the real part of $\eta'$-nucleus optical potential significantly larger than the imaginary part~\cite{NanovaEPJA2018}. 
The $\eta$-nucleon interaction
is observed to be stronger than the $\eta'$-nucleon
interaction~\cite{CzerwinskiPRL2014}. 
In this paper we will focus on the $\eta$-mesic nuclei topic.

Initially, it was predicted that $\eta$-mesic bound states could be formed only with nuclei having a mass number greater than 12, due to the relatively small value of the $\eta$N scattering length estimated in~\cite{BhaleraoPRL1985}. 
However, recent studies of $\eta$ production in photon-~ and hadron- induced reactions, based on different theoretical models, deliver a broad range of the $a_{\eta N}$ scattering length value, from $a_{\eta N}=(0.18,0.16i)$~fm to $a_{\eta N}=(1.03,0.49i)$~fm~\cite{KelkarRPP2013,XiePRC2017,XieEPJA2019}.
The largest values might permit the formation of $\eta$-mesic strongly bound systems even for light nuclei such as $^{4}\hspace{-0.03cm}\mbox{He}$, $^{3}\hspace{-0.03cm}\mbox{He}$, tritium and deuterium~\cite{XiePRC2017,XieEPJA2019,WilkinPRC1993,GreenPRC1996,WycechPRC1995,FixPLB2017,FixPRC2018,BarneaPLB2017,BarneaNPA2017}.

Possible $\eta$-mesic bound states have been searched for in many experiments but, nevertheless, none have found a clear signature of their existence. The measurements allowed us only to observe signals which might be interpreted as indications of the hypothetical $\eta$-mesic nuclei and to determine the upper limits of the total cross section for  bound state formation~\cite{MetagPPNP2017,Sokol2001,BudzanowskiPRC2009,BergerPRL1988,MayerPRC1996,SmyrskiPLB2007,MersmannPRL2007,PapenbrockPLB2014,HaiderJPG2010,AdlarsonPRC2013,AdlarsonNPA2017,SkurzokPLB2018,AdlarsonPLB2020,AdlarsonPRC2020}.

\indent The discovery of the $\eta$-mesic bound systems would be a valuable input towards a better understanding of low-energy meson-nucleon interactions in nuclear media. 
Especially, it would provide information about the properties of the $\eta$ meson inside a nuclear medium as well as being sensitive to the $\eta$-meson's internal structure.
According to the the Quark Meson Coupling model of
Refs.~\cite{BassPLB2006,BassActa2014}, the $\eta$ meson binding energy in nuclear matter is very sensitive to the flavour-singlet component (through $\eta$-$\eta'$ mixing) in the wave function of the $\eta$ meson: a binding energy of 100 MeV is found at nuclear matter density taking the $\eta$-$\eta'$ mixing angle of -20$^{\circ}$, a factor of two larger binding than for a pure octet $\eta$ state.

The investigations of $\eta$-mesic bound states can also be helpful for studying N$^{*}$(1535) resonance properties in medium and to test different theoretical models describing the resonance's internal structure~\cite{InouePRC2002,NagahiroPRC2009,IsgurPRD1978,LiuPRL2016,HyodoPRC2008,GarzonPRC2015,JidoPRC2002}.

\indent This article is aimed at giving an overall status of recent experimental research on $\eta$-mesic bound states, focusing in particular on $\eta$-mesic helium systems.

\section{Status of experimental searches for $\eta$-mesic nuclei}\label{Sec.3}

\begin{figure}[h!]
\centering
\includegraphics[width=5.1cm,height=4.2cm]{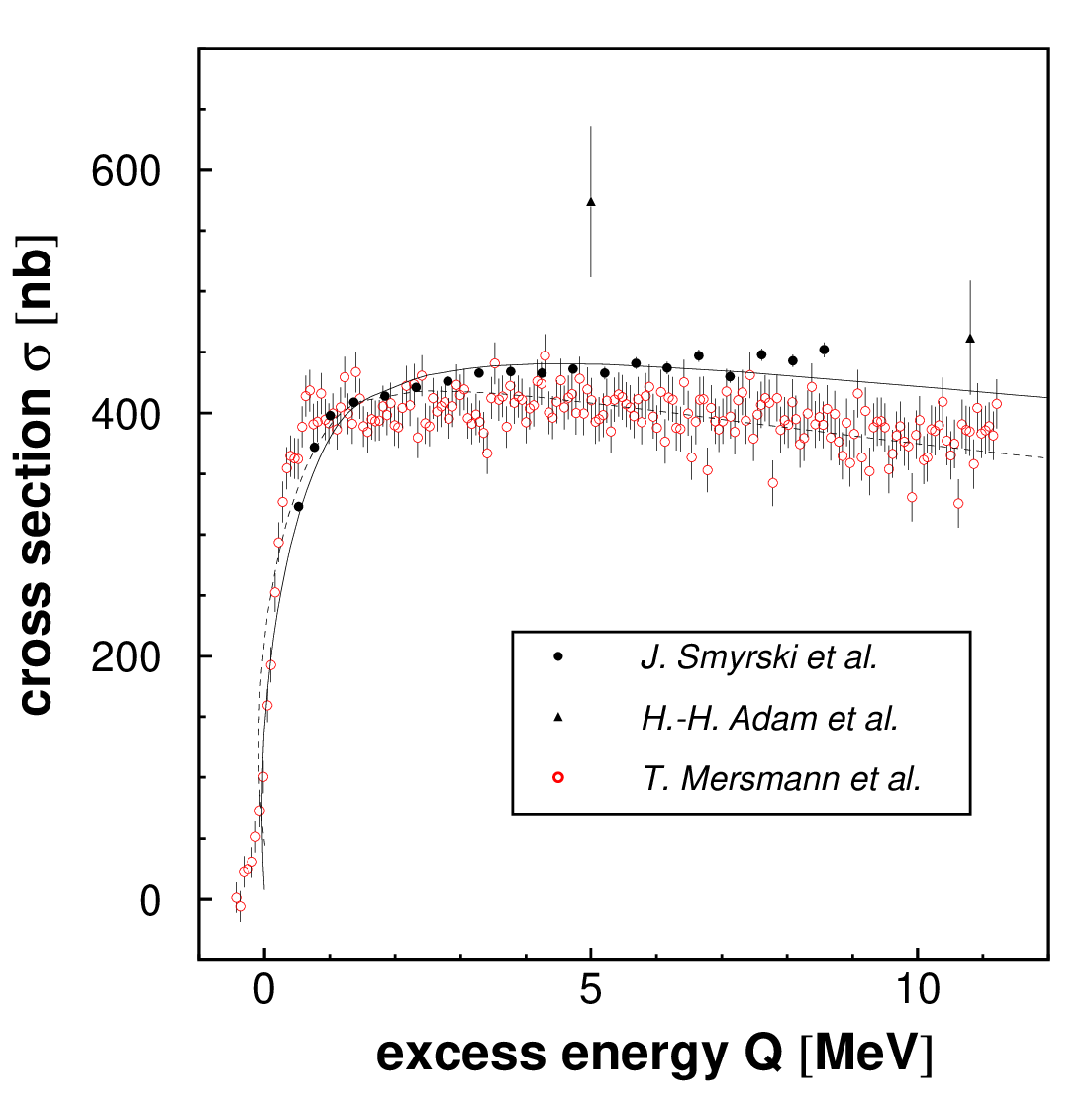} \hspace{0.5cm}
\includegraphics[width=5.0cm,height=4.1cm]{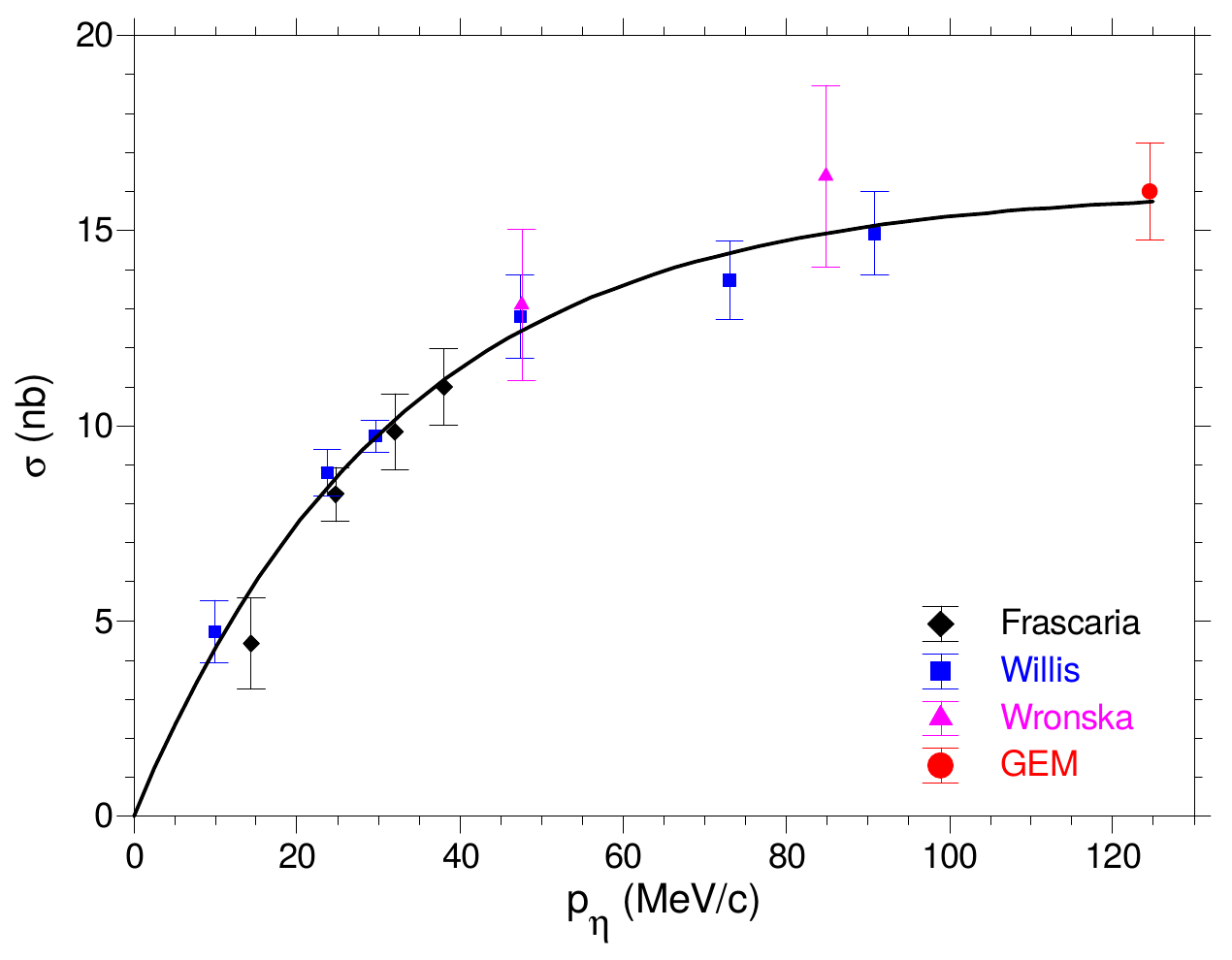}
\caption{(left) Total cross-section for the $dp\rightarrow$ $^{3}\hspace{-0.03cm}\mbox{He}\eta$ reaction measured with the \mbox{COSY-11} facilities (closed circles)~\cite{SmyrskiPLB2007} and (triangles)~\cite{AdamPRC2007} and the \mbox{ANKE} (open circles)~\cite{MersmannPRL2007}. Scattering length fit to the ANKE and COSY-11 data is represented with dashed and solid lines, respectively. (right) Total cross-section for the $dd\rightarrow$ $^{4}\hspace{-0.03cm}\mbox{He}\eta$ reaction as a function of CM momentum obtained from the measurements of Frascaria et al.~\cite{FrascariaPRC1994} (black diamonds), Willis et al.~\cite{WillisPLB1997} (blue squares), Wro{\'n}ska et al.~\cite{WronskaEPJA2005} (magenta triangles) and Budzanowski et al.~\cite{BudzanowskiNPA2009} (red circle). The solid line is to guide the eye.\label{sigma_3He_4He}}
\end{figure}

\indent In parallel to the theoretical investigations, many experiments have been performed to search for $\eta$-mesic bound states, initially in heavy nuclei and currently focusing mainly on light nuclei. The real challenge for experimentalists was the identification of a very weak signal of $\eta$-mesic nuclei over a large background. The measurements for heavy nuclei were carried out with pion~\cite{ChrienPRL1988,JohnsonPRC1993}, photon~\cite{Sokol2001} and hadron beams~\cite{BudzanowskiPRC2009,AfanasievNPB2011}
and are reviewed in Refs.~\cite{KelkarRPP2013,MachnerJPG2015}. In this section we will concentrate on the recent results obtained from experiments devoted to $\eta$-mesic Helium searches. \\
\indent Despite the considerable effort of many theoretical groups in recent years, there are still no model independent calculations which could indicate whether or not $\eta$-mesic helium systems exist. However, there are some experimental observations which may suggest the existence of a bound state, such as a steep rise in the total cross section measured for the \mbox{$dp\rightarrow$ $^{3}\hspace{-0.03cm}\mbox{He}\eta$}~\cite{BergerPRL1988,MayerPRC1996,SmyrskiPLB2007,MersmannPRL2007} and $dd\rightarrow$ $^{4}\hspace{-0.03cm}\mbox{He}\eta$~\cite{FrascariaPRC1994,WillisPLB1997,WronskaEPJA2005} reactions (Fig.~\ref{sigma_3He_4He}), this being a sign of very strong attractive final state interaction (FSI). The FSI and thus the $\eta$-nucleus interaction is much stronger in the case of the $^{3}\hspace{-0.03cm}\mbox{He}\eta$ system indicating that $\eta$ is more likely to bind to $^{3}\hspace{-0.03cm}\mbox{He}$ than to $^{4}\hspace{-0.03cm}\mbox{He}$. 
Another argument in favour of $^{3}\hspace{-0.03cm}\mbox{He}$-$\eta$ bound states is the small value and weak energy dependence of the tensor analysing power $T_{20}$ measured by the ANKE collaboration in the excess energy range Q$\in$(0,11)~MeV which confirms that the very strong variation of the $s$-wave amplitude for $dp\rightarrow$ $^{3}\hspace{-0.03cm}\mbox{He}\eta$ process~\cite{BergerPRL1988,SmyrskiPLB2007,MersmannPRL2007,PapenbrockPLB2014} is associated with an attractive $^{3}\hspace{-0.03cm}\mbox{He}\eta$ interaction. Moreover, the total cross section for $^{3}\hspace{-0.03cm}\mbox{He}\eta$ production is independent of the initial channel. The $^{3}\hspace{-0.03cm}\mbox{He}\eta$ hadron~\cite{SmyrskiPLB2007,MersmannPRL2007} and photo-production~\cite{PfeifferPRL2004,PheronPLB2012} cross sections shows a similar behaviour above threshold which can be assigned to the $^{3}\hspace{-0.03cm}\mbox{He}\eta$ interaction.

The first search for the direct signal of a light $\eta$-nucleus bound state was performed for the $\eta$ photoproduction process $\gamma^{3}$\hspace{-0.03cm}$\mbox{He}\rightarrow \pi^{0}p$X by the TAPS Collaboration~\cite{PfeifferPRL2004}. The difference between the excitation functions measured for two ranges of $\pi^{0}$-proton opening angles of $170^{\circ}-180^{\circ}$ and  $150^{\circ}-170^{\circ}$ in the center-of-mass frame shows enhancement just below the $^{3}\hspace{-0.03cm}\mbox{He}\eta$ threshold. It was interpreted as a possible signature of a {$^{3}\hspace{-0.03cm}\mbox{He}$-$\eta$} bound state where $\eta$ meson captured by one of nucleons inside helium causes excitation of the N$^{*}(1535)$ resonance which then decays into pion-nucleon pair. However, a later experiment carried out with much higher statistics~\cite{PheronPLB2012} showed that the observed structure is an artefact derived from complicated background behaviour.\\
\indent Very promising experiments related to $\eta$-mesic Helium nuclei have been performed at the COSY facility in Forschungszentrum J\"ulich~\cite{WilkinEPJA2017}. The \mbox{COSY-11} group carried out measurements to search for a $\eta$-mesic $^{3}\hspace{-0.03cm}\mbox{He}$ signature in 
the
$dp\rightarrow ppp\pi^{-}$ and $dp\rightarrow$ $^{3}\hspace{-0.03cm}\mbox{He}\pi^{0}$ reactions in the vicinity of the $\eta$ production threshold~\cite{SmyrskiPLB2007,KrzemienIJMPA2009}. The determined excitation functions allowed one to establish the upper limits of the total cross section to be about 270~nb and 70~nb, respectively. 

The search for $^{4}\hspace{-0.03cm}\mbox{He}$-$\eta$ and $^{3}\hspace{-0.03cm}\mbox{He}$-$\eta$ mesic bound systems has been recently extended with the WASA-at-COSY detection setup using a deuteron pellet target and deuteron and proton beams, respectively. The measurements have been performed in three dedicated experiments, in 2008, 2010 and 2014, applying a unique ramped beam technique, namely, continuous and slow changes of the beam momentum around the $\eta$ production threshold in each acceleration cycle allowing for reduced systematic uncertainties with respect to separate runs at fixed beam energies~\cite{SmyrskiPLB2007,AdlarsonPRC2013}.

The $\eta$-mesic bound states were searched for at WASA considering two mechanisms of hypothetical $\eta$-mesic Helium decay: (i) assuming $\eta$ meson absorption on one of the nucleons inside helium, and then its possible propagation in the nucleus via consecutive excitations of the N$^{*}$(1535) resonance, until it decays into the N-$\pi$ pair (e.g. $dd \rightarrow$ $(^{4}\hspace{-0.03cm}\mbox{He}$-$\eta)_{bound} \rightarrow$ N$^{*}$-$^{3}\hspace{-0.03cm}\mbox{He} \rightarrow$ $^{3}\hspace{-0.03cm}\mbox{He} p \pi^{-}$, $pd \rightarrow$ $(^{3}\hspace{-0.03cm}\mbox{He}$-$\eta)_{bound} \rightarrow$ N$^{*}$-$d \rightarrow$ $d p \pi^{0}$) 
and (ii) via $\eta$ meson decay while it is still ``orbiting'' around the nucleus (e.g. $pd\rightarrow$ ($^{3}\hspace{-0.03cm}\mbox{He}$-$\eta)_{bound} \rightarrow$ $^{3}\hspace{-0.03cm} \mbox{He} \ 2\gamma$). 

\begin{figure}[h!]
\centering
\includegraphics[width=5.8cm,height=5.0cm]{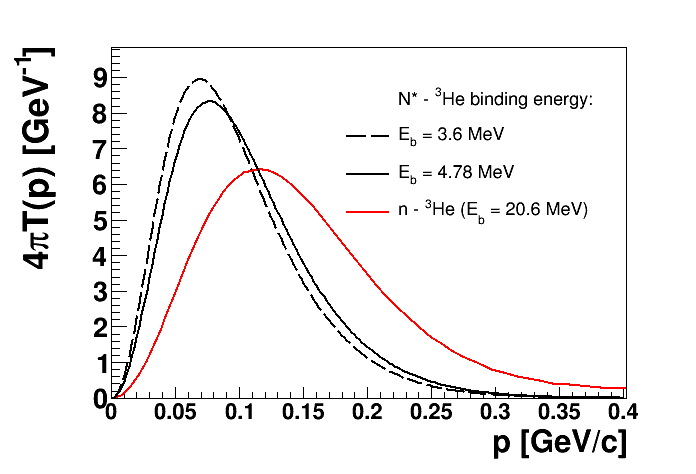}
\includegraphics[width=5.8cm,height=5.0cm]{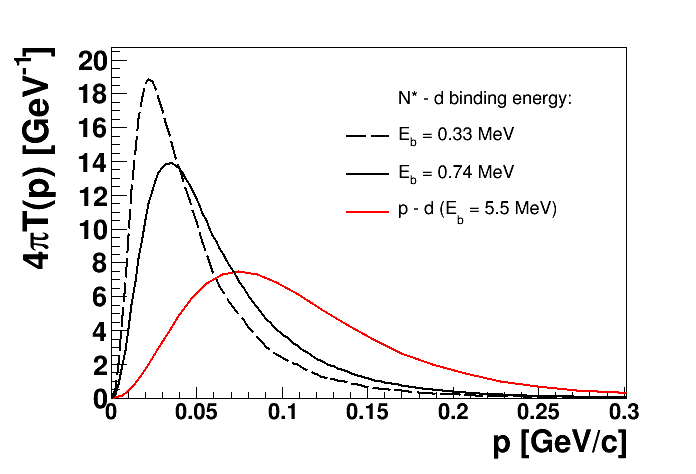}
\caption{ (Left) Momentum distribution of N$^{*}$ (black solid and dashed lines) inside a $^{4}\hspace{-0.03cm}\mbox{He}$ nucleus calculated with N$^{*}$-$^{3}\hspace{-0.03cm}\mbox{He}$ potentials giving 3.6 MeV and 4.78 MeV binding energies and a $n$-$^{3}\hspace{-0.03cm}\mbox{He}$ potential giving a $n$ separation energy of 20.6 MeV (red solid line), respectively. (Right) Momentum distribution of the N$^{*}$ (black solid and dashed lines) inside a $^{3}\hspace{-0.03cm}\mbox{He}$ nucleus calculated with a N$^{*}$-d potential giving 0.74~MeV and 0.33~MeV binding energies and a \mbox{$p$-$^{3}\hspace{-0.03cm}\mbox{He}$} potential giving a $p$ separation energy of 5.5 MeV (red solid line), respectively. The Figures are obtained based on Refs.~\cite{KelkarEPJA2016,KelkarIJMPE2019,KelkarNPA2020}.~\label{Nstar_distr}}  
\end{figure}

\indent In the case of the first mechanism, the kinematics of particles in the final state depends on the momentum of the N$^{*}$ resonance inside the Helium nucleus. 
The first attempt at
determining the N$^{*}$ resonance momentum distribution in the N$^{*}$-$^{3}\hspace{-0.03cm}\mbox{He}$ and \mbox{N$^{*}$-NN} systems has recently been performed and described in Refs.~\cite{KelkarEPJA2016,KelkarIJMPE2019,KelkarNPA2020}. These papers first construct the elementary NN* $\rightarrow$ NN* amplitudes within a $\pi$ plus $\eta$ meson exchange model. The corresponding N*-nucleus potentials are then obtained by folding these amplitudes with the known nuclear density profiles.~The N$^{*}$-$^{3}\hspace{-0.03cm}\mbox{He}$ and N$^{*}$-$d$ momentum distributions are shown in the left and right panels of Fig.~\ref{Nstar_distr}, respectively, for two binding energy values.

%

\noindent They are narrower with respect to the distribution of a neutron in $^{4}\hspace{-0.03cm}\mbox{He}$ or proton in $^{3}\hspace{-0.03cm}\mbox{He}$ which is due to the fact that the N$^{*}$ binding energy is smaller than the energy separation of nucleons in $^{4}\hspace{-0.03cm}\mbox{He}$ and in $^{3}\hspace{-0.03cm}\mbox{He}$. The obtained distributions have been used in Monte Carlo simulations for the recently analysed processes $dd\rightarrow$ ($^{4}\hspace{-0.03cm}\mbox{He}$-$\eta)_{bound}  \rightarrow$ $^{3}\hspace{-0.03cm}\mbox{He} n \pi^{0}$, $dd\rightarrow$ ($^{4}\hspace{-0.03cm}\mbox{He}$-$\eta)_{bound}  \rightarrow$ $^{3}\hspace{-0.03cm}\mbox{He} p \pi^{-}$ and $pd\rightarrow$ ($^{3}\hspace{-0.03cm}\mbox{He}$-$\eta)_{bound}  \rightarrow$ $d p \pi^{0}$~\cite{AdlarsonNPA2017,SkurzokPLB2018,AdlarsonPRC2020}.

\begin{figure}[h!]
\centering
\includegraphics[width=5.8cm,height=4.0cm]{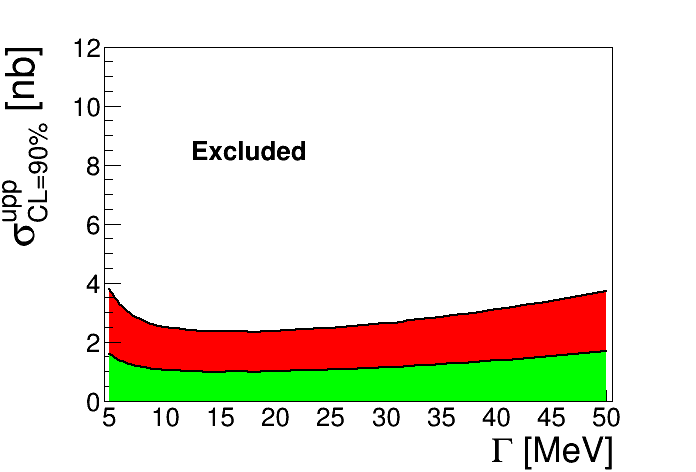}
\includegraphics[width=5.8cm,height=4.0cm]{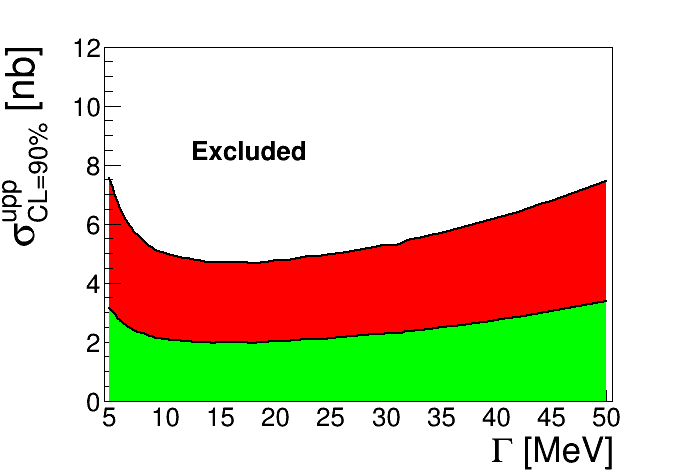}
\caption{Upper limit of the total cross-section for the $dd\rightarrow(^{4}\hspace{-0.03cm}\mbox{He}$-$\eta)_{bound}\rightarrow$ $^{3}\hspace{-0.03cm}\mbox{He} n \pi{}^{0}$ (left panel) and the $dd\rightarrow(^{4}\hspace{-0.03cm}\mbox{He}$-$\eta)_{bound}\rightarrow$ $^{3}\hspace{-0.03cm}\mbox{He} p \pi{}^{-}$ (right panel) reaction as a function of the width of the bound state. The binding energy was fixed to 30~MeV. The upper limit was determined via the simultaneous fit for both channels. The green area denotes the systematic uncertainties. The Figures are adapted from~\cite{AdlarsonNPA2017}.~\label{Result_sigma_upp_both}}  
\end{figure}

\indent The search for $^{4}\hspace{-0.03cm}\mbox{He}$-$\eta$ bound states was carried out by studying the excitation functions for \mbox{$dd\rightarrow$ $^{3}\hspace{-0.03cm}\mbox{He} p \pi^{-}$}~\cite{AdlarsonPRC2013,AdlarsonNPA2017,SkurzokPLB2018} and \mbox{$dd\rightarrow$ $^{3}\hspace{-0.03cm}\mbox{He} n \pi^{0}$}~\cite{AdlarsonNPA2017,SkurzokPLB2018} reactions in vicinity of the $^{4}\hspace{-0.03cm}\mbox{He}\eta$ production threshold (Q$\in(-70,30)$~MeV). The details of the analysis procedures leading to the determination of the excitation functions are described in Refs.~\cite{AdlarsonPRC2013,AdlarsonNPA2017}. The obtained excitation curves do not show any narrow structure below the $\eta$ production threshold, which would be a signature of the bound state. Therefore, the upper limit for the total cross section for the $\eta$-mesic $^{4}\hspace{-0.03cm}\mbox{He}$ formation was determined at 90\% confidence level by fitting the excitation functions with Breit-Wigner function (signal) and with a fixed binding energy and width combined with a second order polynomial (describing the background). For the 2010 data set~\cite{AdlarsonNPA2017} the fit was performed simultaneously for the \mbox{$dd\rightarrow$ $^{3}\hspace{-0.03cm}\mbox{He} p \pi^{-}$} and \mbox{$dd\rightarrow$ $^{3}\hspace{-0.03cm}\mbox{He} n \pi^{0}$} channels taking into account the isospin relation between $n \pi{}^{0}$ and $p \pi{}^{-}$ pairs. The analysis allowed for the first time to determine experimentally the upper limit of the total cross section for the $dd\rightarrow(^{4}\hspace{-0.03cm}\mbox{He}$-$\eta)_{bound}\rightarrow$ $^{3}\hspace{-0.03cm}\mbox{He} n \pi{}^{0}$ process, which varies in the range from 2.5 to 3.5~nb. 
For the
$dd\rightarrow(^{4}\hspace{-0.03cm}\mbox{He}$-$\eta)_{bound}\rightarrow$ $^{3}\hspace{-0.03cm}\mbox{He} p \pi^{-}$ reaction, sensitivity of the cross section of about 6~nb~\cite{AdlarsonNPA2017} was achieved, which is about four times better in comparison with the result obtained in the previous experiment~\cite{AdlarsonPRC2013}. 
The obtained upper limits as a function of the bound state width are presented for both of the studied reactions in Fig.~\ref{Result_sigma_upp_both}.

\begin{figure}[h!]
\centering
\includegraphics[width=8.5cm,height=5.5cm]{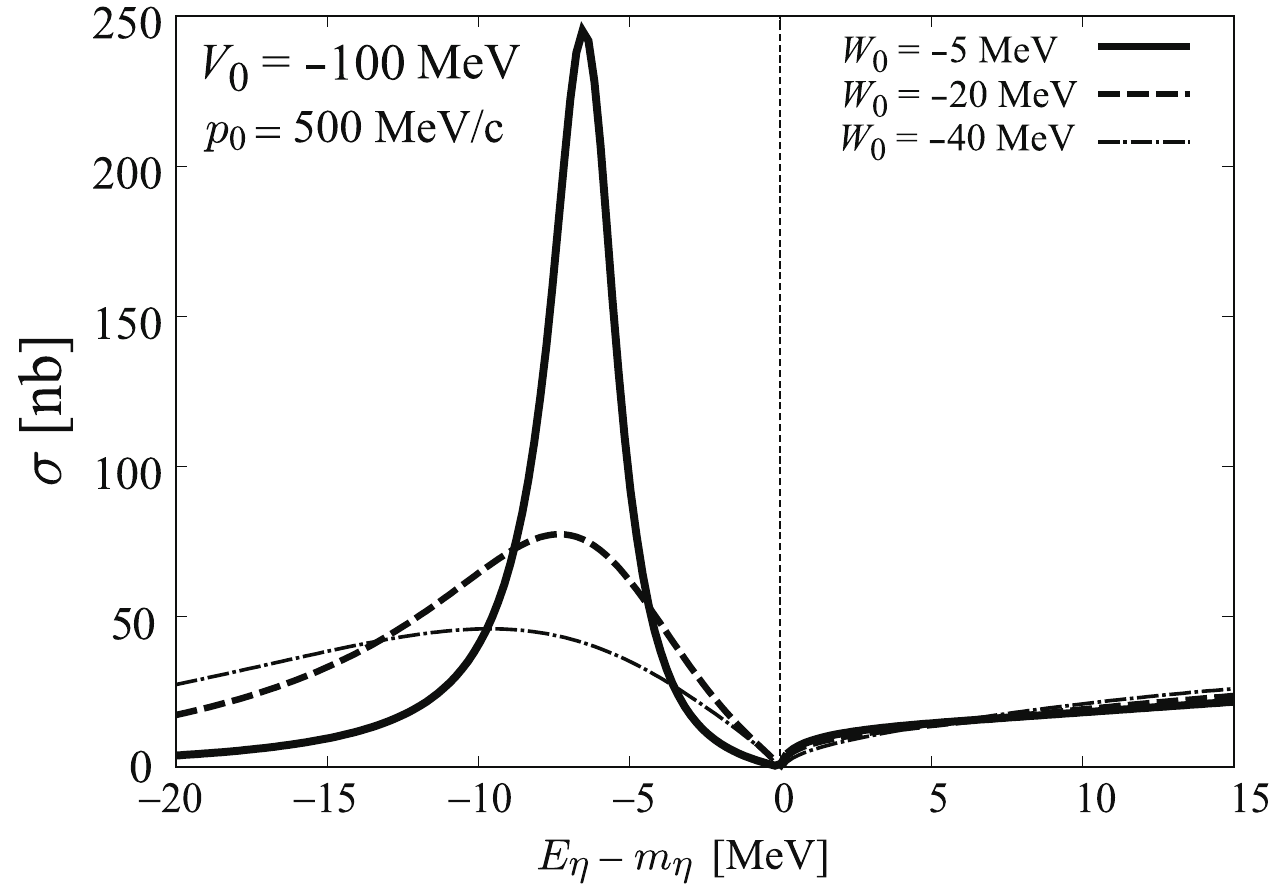}
\caption{Calculated total cross section of the $dd\rightarrow$ ($^{4}\hspace{-0.03cm}\mbox{He}$-$\eta)_{bound}  \rightarrow$ $^{3}\hspace{-0.03cm}\mbox{He}$N$\pi$ reaction for the formation of the $^{4}\hspace{-0.03cm}\mbox{He}$-$\eta$ bound system plotted as function of the excess energy $E_{\eta}-m_{\eta}$ for $\eta$-$^{4}\hspace{-0.03cm}\mbox{He}$ optical potential parameters ($V_{0},W_{0}$)=$-$(100,5), $-$(100,20), $-$(100,40)~MeV (solid, dashed and dotted lines, respectively). The Figure is adapted from Ref.~\cite{IkenoEPJA2017}.~\label{Ikeno}}  
\end{figure}

Due to a lack of theoretical predictions for the $dd\rightarrow(^{4}\hspace{-0.03cm}\mbox{He}$-$\eta)_{bound}\rightarrow$ $^{3}\hspace{-0.03cm}\mbox{He}$N$\pi$ reaction cross sections below the $\eta$ production threshold, in the previous data analyses the bound state signal was assumed to have a Breit-Wigner shape (with fixed binding energy and width)~\cite{AdlarsonNPA2017,AdlarsonPRC2013}. However, recently,  phenomenological calculations of the cross sections in the excess energy range relevant to the $\eta$-mesic nuclear search were presented by Ikeno et al. in Ref.~\cite{IkenoEPJA2017}. These authors provided for the first time the shapes and values of the cross sections for the $dd \rightarrow$ $(^{4}\hspace{-0.03cm}\mbox{He}$-$\eta)_{bound} \rightarrow$ $^{3}\hspace{-0.03cm}\mbox{He}$N$\pi$ process in the excess energy range relevant to the $\eta$-mesic nuclei search. The developed phenomenological model, reproducing the data quite well for the $dd \rightarrow$ $^{4}\hspace{-0.03cm}\mbox{He} \eta$ reaction, allows one to determine the total cross sections for a broad range of $^{4}\hspace{-0.03cm}\mbox{He}$-$\eta$ optical potential parameters ($V_{0}$,$W_{0}$). An example of the calculated total cross section for three different sets of the optical potential parameters is presented in Fig.~\ref{Ikeno}. 
(The contour plot of the determined conversion cross section in the $V_{0}$,$W_{0}$ plane is shown in Fig.~21 of Ref.~\cite{IkenoEPJA2017}.) As a comparison, previous calculations based on an approximation of the scattering amplitude for two body processes~\cite{WycechAPPB2014} allowed one to estimate the cross section for $dd \rightarrow$ $(^{4}\hspace{-0.03cm}\mbox{He}$-$\eta)_{bound} \rightarrow$ $^{3}\hspace{-0.03cm}\mbox{He} p \pi^{-}$ process to be $\sigma\approx$~4.5~nb. 
In the case of $pd \rightarrow$ $(^{3}\hspace{-0.03cm}\mbox{He}$-$\eta)_{bound} \rightarrow$XN$\pi$, only a rough estimation of the total cross section was performed based on the hypothesis that the cross section for the formation of the bound state below the threshold is to first order the same as the $\eta$ meson production cross section close to threshold. It amounts to about 80~nb.

\begin{figure}[h!]
\centering
\includegraphics[width=9.0cm,height=5.5cm]{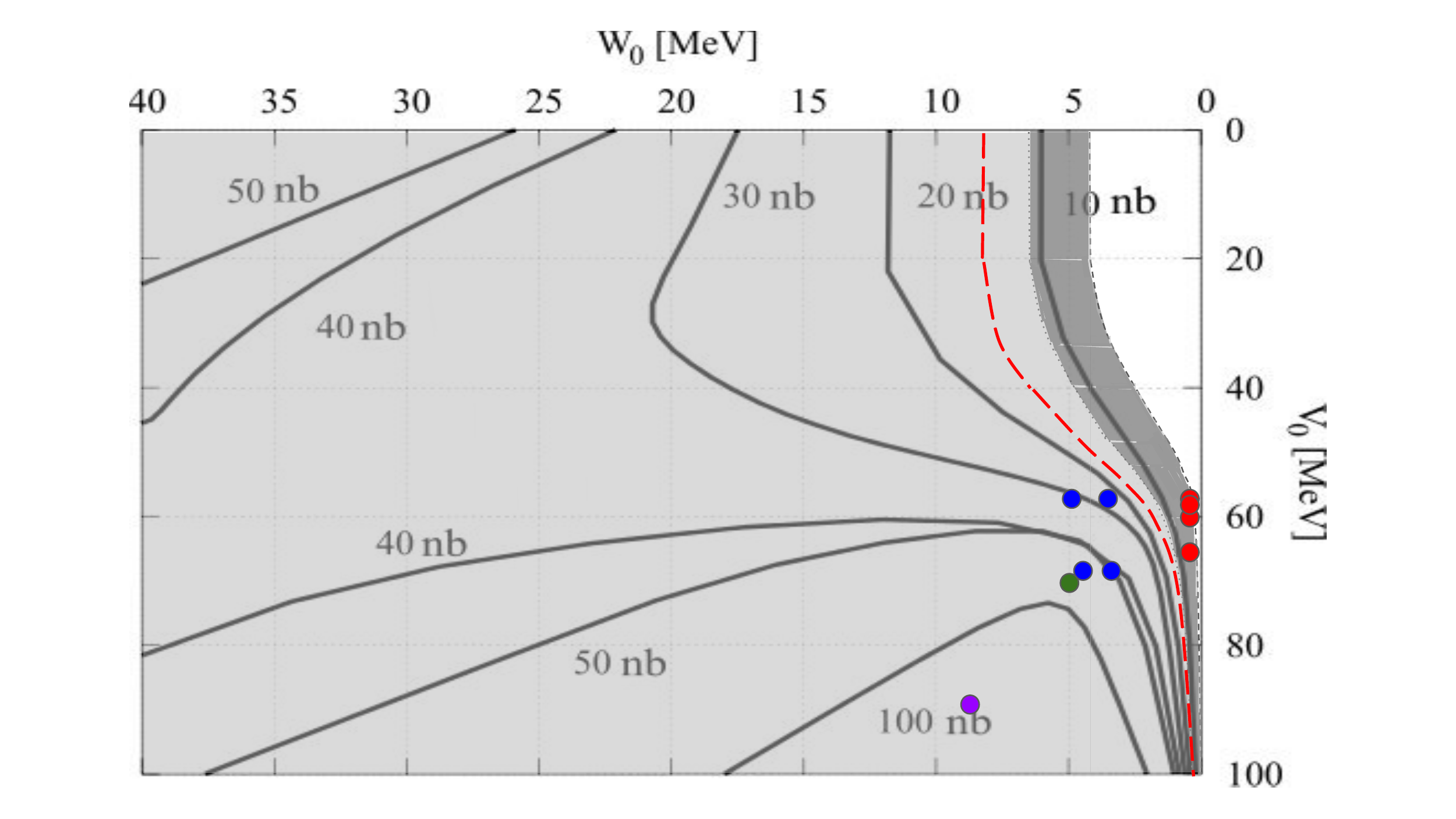}
\caption{Contour plot of the  theoretically determined conversion cross section in the $V_{0} - W_{0}$ plane~\cite{IkenoEPJA2017}. The light shaded area shows the region excluded by the analysis, while the dark shaded area denotes the systematic uncertainty of the $\sigma^{CL=90\%}_{upp}$. The red line extends the allowed region based on a new estimate of errors (see text for details). Dots denote the optical potential parameters corresponding to the predicted $\eta$-mesic $^4$He states. The Figure is adapted from Ref.~\cite{SkurzokPLB2018}.~\label{contour}}
\end{figure}

Fitting the theoretical spectra (convoluted with the experimental resolution of the excess energy) to experimental excitation functions~\cite{AdlarsonNPA2017}, the upper limit of the total cross section (CL=90\%) for creation of $\eta$-mesic nuclei in the \mbox{$dd\rightarrow$ $^{3}\hspace{-0.03cm}\mbox{He}$N$\pi$} reaction was found to vary from about 5.2 nb to about 7.5 nb~\cite{SkurzokPLB2018}. Comparison of the experimentally determined upper limits with the cross sections obtained in Ref.~\cite{IkenoEPJA2017} allowed one to put a constraint on the $\eta$-$^4$He optical potential parameters.
As shown in Fig.~\ref{contour}, 
only extremely narrow and loosely bound states are allowed within the model. Details of the performed studies are presented in Ref.~\cite{SkurzokPLB2018}.

\begin{figure}[h!]
\centering
\includegraphics[width=7.5cm,height=5.5cm]{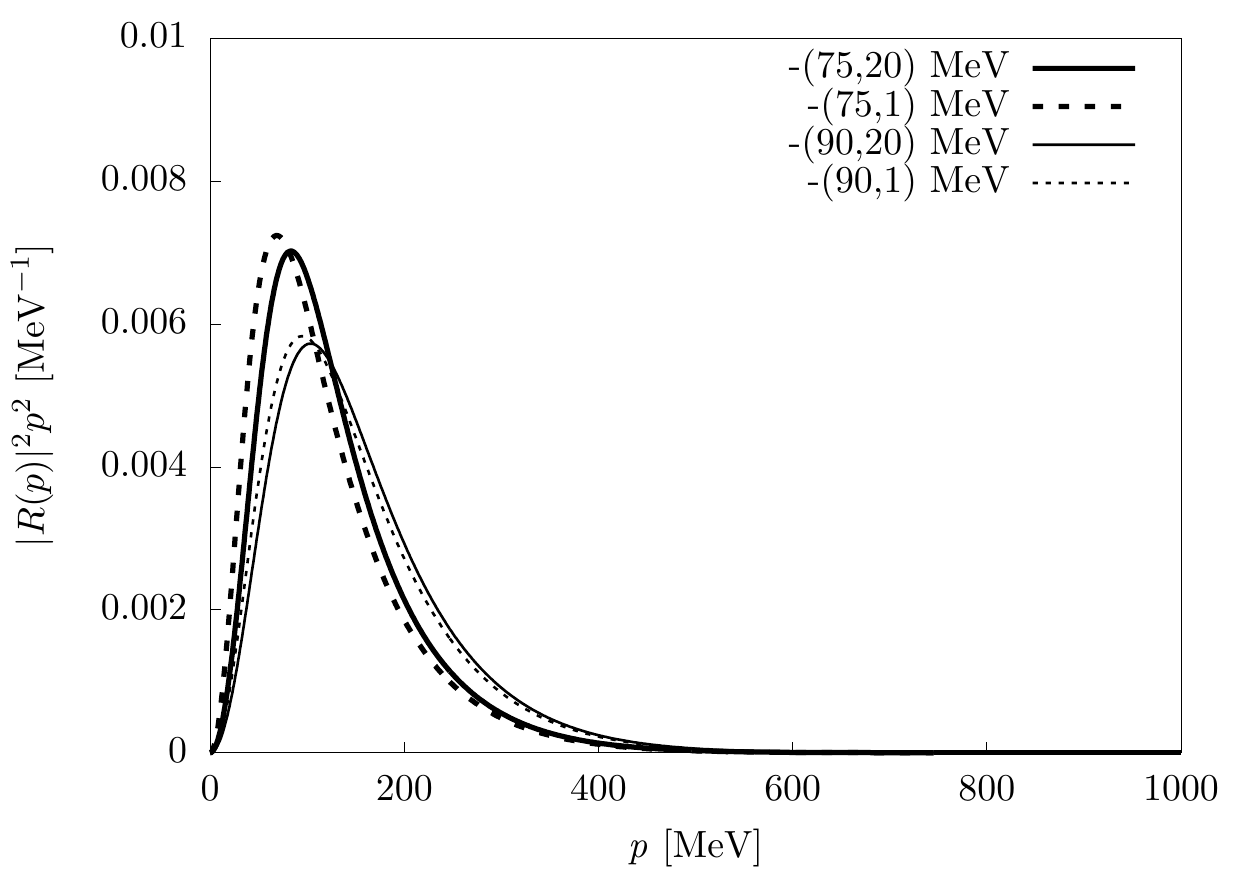}
\caption{Momentum distribution of the $\eta$ meson in $^{3}\hspace{-0.03cm}\mbox{He}$-$\eta$ bound system estimated for ($V_{0},W_{0}$)=$-$(75,20)~MeV (thick solid line), ($V_{0},W_{0}$)=$-$(75,1)~MeV (thick dotted line), ($V_{0},W_{0}$)=$-$(90,20)~MeV (thin solid line), and ($V_{0},W_{0}$)=$-$(90,1)~MeV (thin dotted line). The distributions are normalized to be $1$ in the whole momentum range. The Figure is adapted from Ref.~\cite{SkurzokNPA2020}.~\label{eta_distr}}  
\end{figure}

The last high statistics WASA-at-COSY experimental run (2014) allowed one to search for $\eta$-mesic $^{3}\hspace{-0.03cm}\mbox{He}$ in $pd\rightarrow$ $^{3}\hspace{-0.03cm}\mbox{He}2\gamma$ and $pd\rightarrow$ $^{3}\hspace{-0.03cm}\mbox{He}6\gamma$ processes considering the hypothesis 
of 
mesonic bound state decay into $^{3}\hspace{-0.03cm}\mbox{He} 2\gamma (6\gamma)$ channels.
The reactions have been investigated for the first time using the recently developed theoretical model~\cite{SkurzokNPA2020}
where the bound state is described as a solution of the Klein-Gordon equation. 
The calculations provided relative $^{3}\hspace{-0.03cm}\mbox{He}$-$\eta$ momentum distributions in the $^{3}\hspace{-0.03cm}\mbox{He}$-$\eta$ bound state as well as in-medium branching ratios of $\eta\rightarrow 2\gamma$ and $\eta\rightarrow 3\pi^0$ for different combinations of optical potential parameters. 
The momentum distributions determined for different sets of ($V_{0},W_{0}$) are shown in Fig.~\ref{eta_distr}. The estimated in-medium branching ratios vary from about 2$\cdot$10$^{-5}$ to 7$\cdot$10$^{-4}$ depending on optical potential parameters. Results obtained in the frame of this work were crucial for the Monte Carlo simulations of the considered processes.

\begin{figure}[h!]
\centering
\includegraphics[width=8.0cm,height=6.0cm]{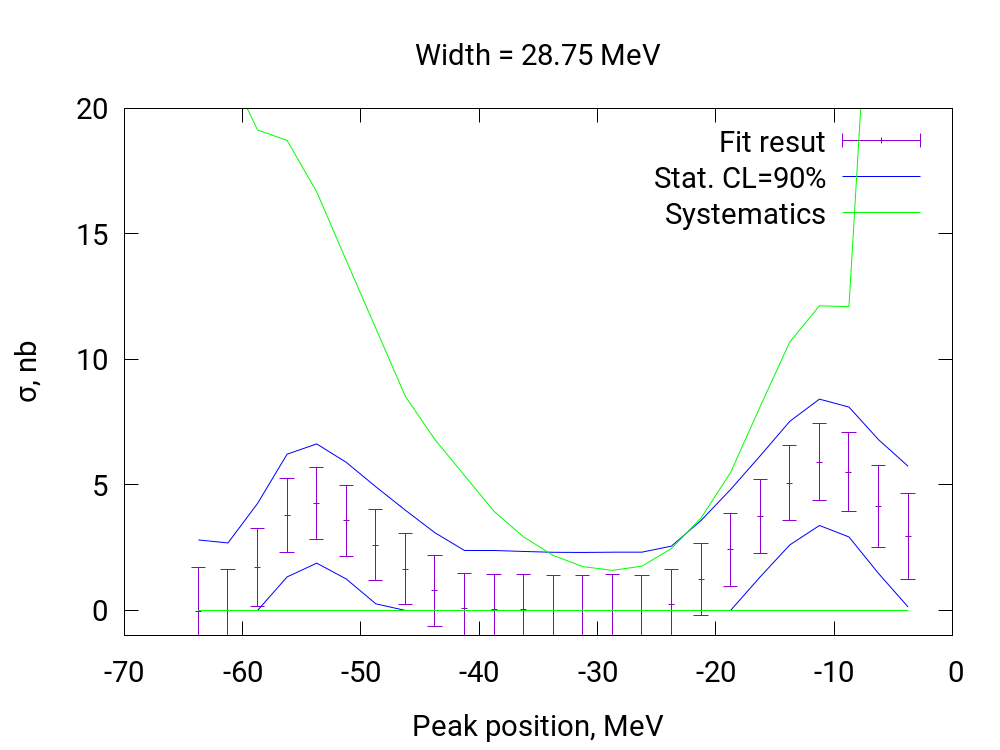}
\caption{Upper limits for the bound state production cross section via $pd\rightarrow$ ($^{3}\hspace{-0.03cm}\mbox{He}$-$\eta)_{bound} \rightarrow$ $^{3}\hspace{-0.03cm}\mbox{He}$($\eta~ \mbox{decays})$ as function of the peak position for fixed width $\rm \Gamma$=28.75~MeV. The values of the Breit-Wigner amplitude $\sigma$ are shown with statistical uncertainties. The range of possible bound state production cross section obtained based on statistical uncertainty corresponding to $90\%$ confidence level is shown by blue lines. The range of possible bound state production cross section including  systematic uncertainty is shown by green lines. Figure is adapted from Ref.~\cite{AdlarsonPLB2020}.}	\label{upper_limit_1d}
\end{figure}

The obtained excitation functions for both measured channels show a slight hint of the signal of a possible bound state for a width greater than 20~MeV and binding energy in the range from 0 to 15~MeV. However, the observed indication is within the systematic error which does not allow one to conclude whether or not the bound state is created by the considered mechanism. Therefore, finally, the upper limit of the total cross section at the CL=90\% was determined for the $\eta$-mesic $^{3}\hspace{-0.03cm}\mbox{He}$ nucleus creation followed by the $\eta$ meson decay, by fitting simultenously the excitation functions for both reactions with a Breit-Wigner +  polynomial (signal+background) taking into account the branching ratio relation between $\eta \rightarrow 2\gamma$ and $\eta \rightarrow 3\pi^{0}$ in vacuum. The estimated upper limit varies between 2 nb to 15 nb depending on the bound state parameters (binding energy, width)~\cite{AdlarsonPLB2020}. It is shown in Fig.~\ref{upper_limit_1d} for the bound state width $\rm \Gamma$=28.75~MeV. The determined upper limit is much lower than the limit obtained in~\cite{KrzemienIJMPA2009} for $pd \rightarrow$ $(^{3}\hspace{-0.03cm}\mbox{He}-\eta)_{bound}$ $\rightarrow$ $^{3}\hspace{-0.03cm}\mbox{He} \pi^{0}$ (70~nb) and 
is comparable with upper limits obtained in~\cite{AdlarsonNPA2017,SkurzokPLB2018}.\\

\begin{figure}[h!]
\centering
\includegraphics[width=8.5cm,height=6.0cm]{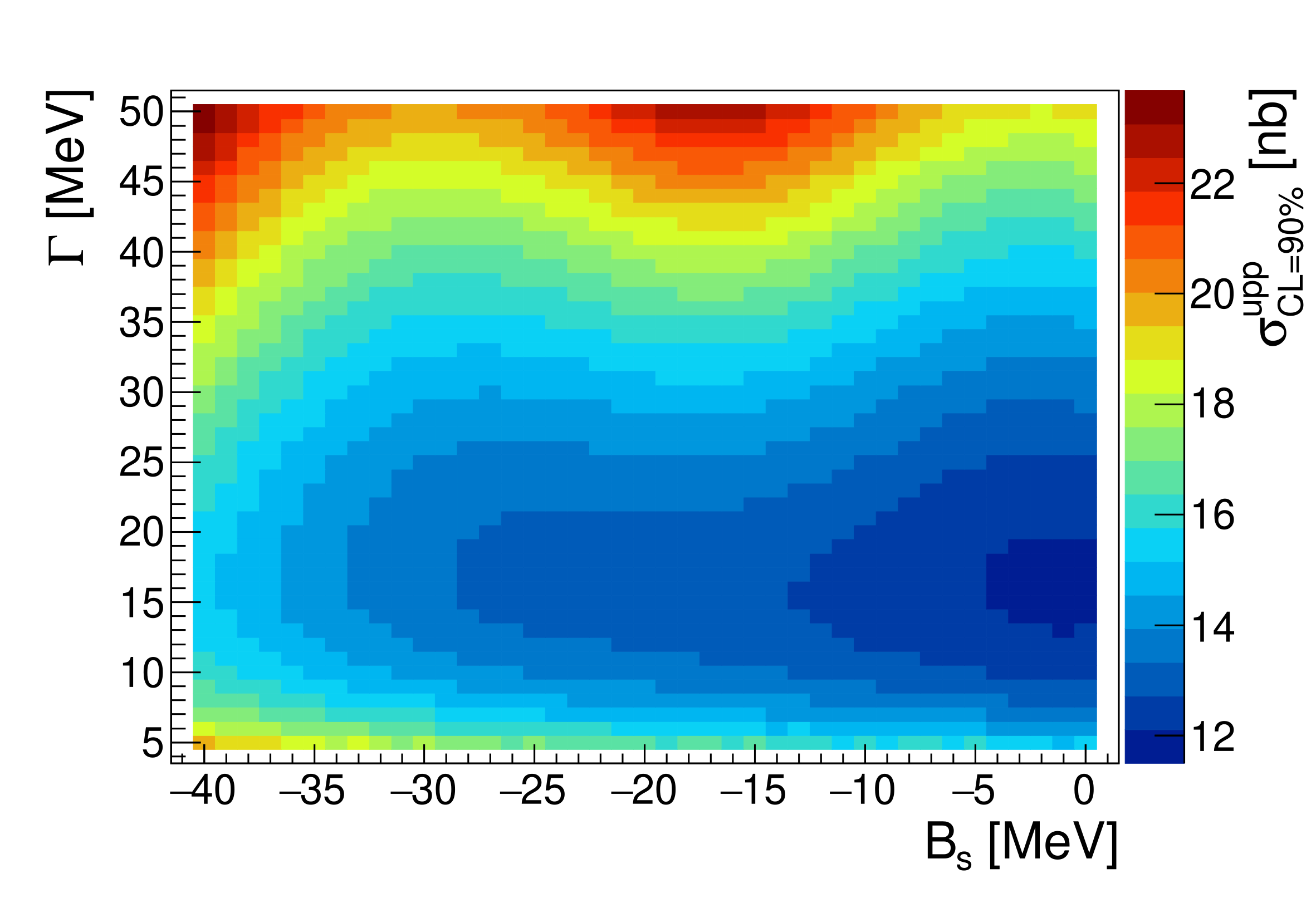}
\caption{The upper limit of the total cross section at 90\% confidence level obtained for \mbox{$pd\rightarrow$ $(^{3}\hspace{-0.03cm}\mbox{He}$-$\eta)_{bound} \rightarrow$ $d p \pi^{0}$} process assuming different bound state parameters, $B_{s}$ and $\Gamma$. The Figure is adapted from Ref.~\cite{AdlarsonPRC2020}.}\label{upper_limit_2d}
\end{figure}

Recently the data analysis for $pd \rightarrow d p \pi^{0}$ reaction corresponding to mechanism (i) has been completed~\cite{AdlarsonPRC2020}. Since the narrow resonance-like structure associated with an $\eta$-mesic $^{3}\hspace{-0.03cm}\mbox{He}$ nuclei was not observed, the upper limit of the total cross section for the \mbox{$pd\rightarrow$ $(^{3}\hspace{-0.03cm}\mbox{He}$-$\eta)_{bound} \rightarrow$ $d p \pi^{0}$} process was estimated at the CL 90\%. The upper limit varies from 13 to 24 nb for the bound state parameters \mbox{$B_{s}\in(0,40)$ MeV} and \mbox{$\Gamma\in(0,50)$ MeV} as it is shown in Fig.~\ref{upper_limit_2d}. The determined limits do not exclude the bound states predicted with $\eta$-nucleon scattering lengths with a real part of about 1~fm. 

\section{Summary and Perspectives}

In this report recent experimental results concerning $\eta$-mesic bound states have been discussed. We have focused mostly on the $\eta$-mesic Helium searches presenting the latest experimental data analyses with application of current phenomenological models. The performed measurements result in the valuable upper limits of the total cross sections for ($^{4}\hspace{-0.03cm}\mbox{He}$-$\eta)_{bound}$ and ($^{3}\hspace{-0.03cm}\mbox{He}$-$\eta)_{bound}$ production and decay considering different mechanisms. A new analysis is currently being carried out to simultaneously adjust the excitation function for processes running according to two different mechanisms






\begin{acknowledgements}
I acknowledge the support from the Polish National Science Center through grant No. 2016/23/B/ST2/00784.

\end{acknowledgements}


\end{document}